# Transformer networks for continuous gravitational-wave searches

Prasanna. M. Joshi* and Reinhard Prix
*Max Planck Institute for Gravitational Physics (Albert-Einstein-Institute), 30167 Hannover, Germany*
*and Leibniz Universität Hannover, 30167 Hannover, Germany*

Wide-parameter-space searches for continuous gravitational waves using semicoherent matched-filter methods require enormous computing power, which limits their achievable sensitivity. Here we explore an alternative search method based on training neural networks as classifiers on detector strain data with minimal preprocessing. Contrary to previous studies using convolutional neural networks (CNNs), we investigate the suitability of the transformer architecture, specifically the vision transformer (ViT). We establish sensitivity benchmarks using the matched-filter $\mathcal{F}$-statistic for ten targeted searches over a ten day timespan, and ten directed and six all-sky searches over a one day timespan. We train ViTs on each of these benchmark cases. The trained ViTs achieve essentially matched-filter sensitivity on the targeted benchmarks, and approach the $\mathcal{F}$-statistic detection probability of $p_{\mathrm{det}} = 90\%$ on the directed ($p_{\mathrm{det}} \approx 85$–$89\%$) and all-sky benchmarks ($p_{\mathrm{det}} \approx 78$–$88\%$). Unlike the CNNs in our previous studies, which required extensive manual design and hyperparameter tuning, the ViT achieves better performance with a standard architecture and minimal tuning.

## I. INTRODUCTION

Continuous gravitational waves (CWs) are long-lasting, quasi-monochromatic gravitational waves emitted by rapidly spinning neutron stars with a nonaxisymmetric deformation. Many of these signals are expected to fall in the sensitive band of ground-based gravitational wave detectors: Advanced LIGO [1], Advanced VIRGO [2], and KAGRA [3]. Owing to their small amplitude, it is necessary to analyze a long duration of data in order to be able to find CWs in the detector strain.

The theoretically most sensitive search method, known as coherent matched filtering, involves cross-correlating the detector strain data with signal waveform templates. When considering a wide parameter space of possible signals, using a coherent data timespan of several months would result in an infeasible computing cost due to the large number of required signal templates, rendering this approach unusable. The state-of-the-art method is semicoherent matched filtering, which splits the total data duration into shorter segments, computing the coherent matched filter over each segment, and then combining their results incoherently (i.e., by summing per-segment power rather than complex amplitudes). The parameters of this search method (e.g., the number and length of segments, template-bank density, etc.) can be tuned to get maximum sensitivity (for this method) at a given computational cost. However, such methods still require massive computational power in order to achieve good sensitivity to CW signals. For a more complete description of the different search methods and searches based on matched filtering, see recent reviews on the topic, e.g., [4,5] and references therein.

Here we focus instead on deep learning as an alternative method to lower the computational cost and potentially improve the sensitivity of CW searches. There have been a number of studies already exploring the potential of deep neural networks to improve different aspects of CW searches: their application to clustering of search candidates was explored in [6,7], and [8,9] used them to improve follow-up searches. A full search pipeline using neural networks was developed in [10,11], and studies on how to mitigate the effect of instrumental noise artifacts on neural-network sensitivity have shown encouraging results [12,13]. Various search pipelines involving neural networks [mostly convolutional neural networks (CNNs)] were featured in [14]. Some neural-network-based search methods have also been developed for transient CWs in recent years [15–17].

This work is in the line of [18–21], where the CW search is formulated as an image-classification problem directly on the detector strain data (transformed in the time-frequency domain), and deep neural networks are trained to perform this classification. These studies were based on CNNs, which used to be the best image-classification architecture. However, as we discussed in [20], the default small-kernel CNNs turn out to be ill suited for the nature of the CW search problem, which is why they are struggling with weak signals spread out in frequency and time and buried deep in the noise. We further showed that a large-kernel CNN purposefully designed for the signal characteristics *can* reach (close to) optimal matched-filter sensitivity on a targeted-search benchmark for a ten-day timespan. Using these principles we were also able to improve the CNN sensitivity on ten-day all-sky and directed search benchmarks [21], but still falling short of full matched-filter sensitivity.

*Contact author: prasanna.mohan.joshi@aei.mpg.de

Here we take a step back and investigate the suitability of a completely different family of neural-network architectures: transformers [22]. In particular, the vision transformer (ViT) [23] is now rivaling and surpassing CNNs as the state-of-the-art architecture for various image tasks such as image classification, object detection, etc. [23–27]. We therefore study the efficacy of a ViT-based architecture trained for targeted, directed, and all-sky CW searches. Note that we initially considered the ViT performance on the ten-day targeted search case of [20] and found that an essentially "off-the-shelf" ViT can achieve (close to) matched-filtering performance (see Sec. IV A) without any special architecture tuning, contrary to CNNs. This led us to extend the scope to wide-parameter-space searches (directed and all-sky), but for this initial study limited to the easier one-day timespan benchmarks similar to those used in [18,19]. Training times on the longer-timespan wide-parameter-space searches (such as ten days) are substantially longer and are postponed to future work. As in previous works, we compare the neural-network sensitivity to the corresponding (near-optimal) $\mathcal{F}$-statistic matched-filter search method (defining our *benchmarks*), and we test how its sensitivity depends on various CW signal parameters such as amplitude, frequency, and sky position.

This paper is organized as follows: in Sec. II, we introduce the CW search benchmarks; in Sec. III, we describe the transformer architecture, training process, and computation of metrics; in Sec. IV, we present the comparison of the sensitivity of the ViT search to the matched filter search and comment on the generalization properties of the trained ViTs; and we discuss the conclusions of our work in Sec. V.

## II. CW SEARCH BENCHMARKS

In this section we define the CW search benchmarks used to characterize the performance of the ViT-based search method relative to a (near-optimal) matched-filter search using the coherent $\mathcal{F}$-statistic [28]. For each benchmark case described in the following, we estimate the corresponding matched-filter sensitivity, which defines the (close to) best achievable result on any given search challenge.

TABLE I. Parameters defining the benchmark search cases used here. Detectors H1 and L1 refer to the LIGO Hanford and Livingston interferometers, respectively. The sky-position parameters $(\alpha, \delta)$ refer to longitude and latitude in equatorial coordinates.

| Targeted search | |
|---|---|
| Start time | 1 200 300 463 s |
| Duration | 10 days |
| Detectors | H1 and L1 |
| Noise | Stationary, white, Gaussian |
| Frequency $f(\tau_{\mathrm{ref}})$ | 20, 100, 200, 500, 1000 Hz |
| Spindown $\dot{f}(\tau_{\mathrm{ref}})$ | $-10^{-10}$ Hz s$^{-1}$ |
| $\tau_{\mathrm{ref}}$ | 1 200 732 463 s |
| Sky position $(\alpha, \delta)$ | Sky-A: (6.123771, 1.026457) rad |
| | Sky-B: (2.119314, 0.299076) rad |

| Directed search | |
|---|---|
| Start time | 1 200 300 463 s |
| Duration | 1 day |
| Detectors | H1 and L1 |
| Noise | Stationary, white, Gaussian |
| $\tau_{\mathrm{ref}}$ | 1 200 343 663 s |
| Sky-position $(\alpha, \delta)$ | G347: (4.509371, −0.695189) rad |
| | CasA: (6.123771, 1.026457) rad |
| Reference Frequency $f_{\mathrm{ref}}$ | 20, 100, 200, 500, 1000 Hz |
| Frequency range | $f \in [f_{\mathrm{ref}}, f_{\mathrm{ref}} + 50\ \mathrm{mHz}]$ |
| Spin-down range | $\dot{f} \in [-f/\tau, 0]$ Hz s$^{-1}$ |
| Second order spin-down | $\ddot{f} \in [0, 5f/\tau^2]$ Hz s$^{-2}$ |
| Characteristic age $(\tau)$ | G347: 1600 yr |
| | CasA: 330 yr |

| All-sky search | |
|---|---|
| Start time | 1 200 300 463 s |
| Duration | 1 day |
| Detectors | H1 and L1 |
| Noise | Stationary, white, Gaussian |
| $\tau_{\mathrm{ref}}$ | 1 200 343 663 s |
| Sky-region | All-sky |
| Reference Frequency $f_{\mathrm{ref}}$ | 20, 100, 200, 500, 1000 Hz |
| Frequency range | $f \in [f_{\mathrm{ref}}, f_{\mathrm{ref}} + 50\ \mathrm{mHz}]$ |
| Spin-down range | $\dot{f} \in [-10^{-10}, 0]$ Hz s$^{-1}$ |

### A. Benchmark definition

The details for all benchmarks are given in Table I. We consider two targeted searches spanning ten days, and three wide-parameter-space searches (two directed and one all-sky) spanning one day, in order to keep training times manageable for this initial exploration of ViTs, as mentioned in the Introduction. Each search is assuming two detectors (H1 and L1), stationary white noise, and is considered at five different "reference" frequencies $f_{\mathrm{ref}}$, namely 20 Hz, 100 Hz, 200 Hz, 500 Hz, and 1000 Hz.

The targeted searches are for two different sky positions, Sky-A and Sky-B. Sky position Sky-A, which corresponds to the supernova remnant Cassiopeia A (CasA), was chosen because the corresponding signal has a small bandwidth over the given timespan, which is typically easier for neural networks to learn [20,21], while sky position Sky-B corresponds to the signal with the widest bandwidth in the search timespan, making it potentially the hardest sky position. These targeted benchmarks are the same as those used in [20].

The two directed search benchmarks consider signals from the supernova remnants CasA and G347.3-0.5 (G347), respectively. Each directed search case is defined for a bandwidth of 50 mHz at the five reference frequencies $f_{\mathrm{ref}}$, and the ranges of first- and second-order frequency derivatives are given as a function of frequency and the characteristic age of the corresponding supernova remnant.

The all-sky search benchmark cases are similarly defined for a bandwidth of 50 mHz at the five reference frequencies, but with a fixed range in first-order spindown.

Note that the wide-parameter-space benchmarks are similar but not identical to those used in [19], which had a timespan of $10^5\ \mathrm{s} \approx 1.16$ days, while we chose to use more "canonical" spans of integer multiples of days instead.

### B. Sensitivity estimation

The sensitivity of a search can be characterized by the detection probability ($p_{\mathrm{det}}$) at fixed false-alarm probability ($p_{\mathrm{fa}}$) on a population of signals at fixed amplitude $h_0$. The amplitude of a CW signal relative to the noise floor (given by the power spectral density $S_{\mathrm{n}}$) is often conveniently expressed in terms of the sensitivity depth $\mathcal{D}$, defined as

$$\mathcal{D} \equiv \frac{\sqrt{S_{\mathrm{n}}}}{h_0}. \tag{1}$$

We can therefore represent the sensitivity of a search method independently of the noise floor in terms of the 90%-upper-limit sensitivity depth, denoted as $\mathcal{D}^{90\%}$, which is the signal depth $\mathcal{D}$ for which the search has $p_{\mathrm{det}} = 90\%$ at a given $p_{\mathrm{fa}}$.

The signal power $\rho^2$ (also referred to as squared signal-to-noise ratio for coherent searches) is defined (e.g., see [29]) as

$$\rho^2 \equiv \frac{4}{25}\frac{T_{\mathrm{data}}}{\mathcal{D}^2}R^2(\theta), \tag{2}$$

where $T_{\mathrm{data}}$ is the total duration of data from all the detectors and $R(\theta)$ is a geometric antenna-response factor $\sim\mathcal{O}(1)$ that depends on the signal sky position and polarization angles.

We measure the matched-filter sensitivity of an $\mathcal{F}$-statistic-based search for each of the benchmark cases described above, expressed in terms of the 90% sensitivity depth $\mathcal{D}^{90\%}$ at a false alarm level of $p_{\mathrm{fa}} = 1\%$. For the targeted benchmarks, $\mathcal{D}^{90\%}$ can be easily estimated directly using the approach developed in [29,30].

For the directed and all-sky benchmarks, we use the WEAVE code [31] to perform template-bank searches and measure the resulting sensitivity. The template banks are generated with a mismatch parameter of 0.1, and the resulting numbers of templates $\mathcal{N}_{\mathrm{T}}$ are given in Table II. By running $10^5$ repeated searches on pure Gaussian noise, we obtain the noise distribution of the loudest $\mathcal{F}$-statistic candidate over the search parameter space, which yields the detection threshold $\mathcal{F}_{\mathrm{th}}$ such that $p_{\mathrm{fa}} \equiv P(\mathcal{F} > \mathcal{F}_{\mathrm{th}} | h_0 = 0) = 1\%$ over the 50 mHz bandwidth searched. The resulting thresholds are given in Table II. The corresponding detection probability $p_{\mathrm{det}}$ is computed by performing repeated searches on signals added to Gaussian noise at a constant depth $\mathcal{D}$ using the above thresholds. By varying $\mathcal{D}$ we can therefore find the value $\mathcal{D}^{90\%}$ such that $p_{\mathrm{det}} = 90\%$ at fixed $p_{\mathrm{fa}} = 1\%$, used to characterize the sensitivity of a search. The resulting values of $\mathcal{D}^{90\%}$ for all the benchmarks are given in Table III.[1]

TABLE II. Number of templates $\mathcal{N}_{\mathrm{T}}$ used in the $\mathcal{F}$-statistic WEAVE search and the corresponding $\mathcal{F}$-statistic thresholds $\mathcal{F}_{\mathrm{th}}$, corresponding to a false-alarm level of $p_{\mathrm{fa}} = 1\%$ per 50 mHz bandwidth for each of the directed and all-sky benchmarks.

| | $f_{\mathrm{ref}}$ | 20 Hz | 100 Hz | 200 Hz | 500 Hz | 1000 Hz |
|---|---|---|---|---|---|---|
| G347 | $\mathcal{N}_{\mathrm{T}}$ | $3.2\times10^4$ | $1.4\times10^5$ | $2.9\times10^5$ | $6.9\times10^5$ | $1.4\times10^6$ |
| | $\mathcal{F}_{\mathrm{th}}$ | 34.5 | 37.2 | 38.6 | 40.3 | 41.9 |
| CasA | $\mathcal{N}_{\mathrm{T}}$ | $1.3\times10^5$ | $6.7\times10^5$ | $1.3\times10^6$ | $3.3\times10^6$ | $6.7\times10^6$ |
| | $\mathcal{F}_{\mathrm{th}}$ | 38.2 | 41.6 | 43.1 | 45.1 | 46.6 |
| All-sky | $\mathcal{N}_{\mathrm{T}}$ | $8.9\times10^5$ | $1.3\times10^7$ | $4.7\times10^7$ | $2.8\times10^8$ | $1.1\times10^9$ |
| | $\mathcal{F}_{\mathrm{th}}$ | 41.9 | 47.8 | 50.8 | 54.4 | 57.4 |

TABLE III. Upper-limit sensitivity depths $\mathcal{D}^{90\%}$ at $p_{\mathrm{fa}} = 1\%$ for the $\mathcal{F}$-statistic searches (top) and the ViT searches (bottom) for each of the search benchmarks. Note that for the targeted matched-filter searches, $\mathcal{D}^{90\%}_{\mathcal{F}}$ only depends on the sky position and not on frequency.

| $\mathcal{D}^{90\%}_{\mathcal{F}}\,[/\sqrt{\mathrm{Hz}}]$ | 20 Hz | 100 Hz | 200 Hz | 500 Hz | 1000 Hz |
|---|---|---|---|---|---|
| Sky-A | 86.1 | 86.1 | 86.1 | 86.1 | 86.1 |
| Sky-B | 81.6 | 81.6 | 81.6 | 81.6 | 81.6 |
| G347 | 17.2 | 16.4 | 16.2 | 15.8 | 15.6 |
| CasA | 16.7 | 15.9 | 15.6 | 15.1 | 14.9 |
| All-sky | 14.9 | 14.2 | 13.6 | 13.3 | 12.8 |

| $\mathcal{D}^{90\%}_{\mathrm{ViT}}\,[/\sqrt{\mathrm{Hz}}]$ | 20 Hz | 100 Hz | 200 Hz | 500 Hz | 1000 Hz |
|---|---|---|---|---|---|
| Sky-A | 85.5 | 84.9 | 85.0 | 85.0 | 85.4 |
| Sky-B | 80.2 | 81.0 | 80.1 | 80.2 | 77.3 |
| G347 | 16.3 | 15.9 | 15.6 | 15.1 | 14.7 |
| CasA | 16.4 | 15.6 | 15.3 | 14.5 | 13.9 |
| All-sky | 14.6 | 13.4 | 12.7 | 11.9 | 11.1 |

## III. VISION TRANSFORMER SEARCH

As in previous works in this line of research [19–21,29], we formulate the problem of CW detection in terms of image classification. Contrary to these previous studies based on the CNN architecture, here we explore for the first time the suitability of the vision transformer [23], a variant of the original transformer networks [22] adapted to image classification.

### A. Preparation of input image

The input to the ViT is a two-dimensional multichannel image, which we construct directly from short Fourier transform (SFTs), the standard input data format [32] of many CW search pipelines.

Contrary to our previous works [20,21], where we transformed these standard 1800 s SFTs into longer spectrograms, here we use the SFTs *directly*, simply stacking consecutive SFTs along one image axis, with frequency being the other, and (as before) real and imaginary parts forming two channels per detector (for two detectors our input images therefore have four channels). The image dimensions are therefore the number of SFTs in the search timespan along the time axis and the number of frequency bins in the search bandwidth along the frequency axis.

The search bandwidth is chosen as the bandwidth of the widest signal for every benchmark type, which is the smallest possible input window without truncating any signals, allowing for the fastest training speed. The widest signal bandwidths are 22.2 mHz, 12.2 mHz, and 5.6 mHz for the targeted, directed, and all-sky benchmarks, respectively. This results in corresponding input SFT image dimensions of $480 \times 40$, $48 \times 22$, and $48 \times 10$ pixels, respectively, along the time and frequency axis.

### B. Network architecture

Following [23] the SFT image is divided into two-dimensional patches with fixed dimensions, namely the *patch width* and *patch height* along the time and frequency axes. Each patch is flattened to form a one-dimensional input token for the transformer network.

The patch height is set to twice the widest signal bandwidth over the patch width, and we use half-overlapping patches along the frequency axis, with no overlap along the time axis. This ensures that any signal will be entirely contained within at least one patch along the frequency axis, in accordance with our design principles developed in [20,21].

The patch width therefore determines the total size of the patches, the dimension of the resulting flatted tokens, and the total number of patches (i.e., tokens) the SFT input image is broken into. This is a significant hyperparameter that we empirically optimize for best performance, with the following best values for patch width $\times$ patch height found as $48 \times 14$, $2 \times 4$, and $2 \times 4$ for the targeted, directed, and all-sky ViTs, respectively, resulting in 40, 240, and 96 input tokens.

Similar to [23], a learnable fully connected, linear embedding is applied to each token that maps each token to a latent vector size of 512. A learnable one-dimensional positional embedding is added to the above in order to retain the positional information of each patch.

The ViT contains a chain of four transformer encoders followed by the output block. The structure of the transformer encoder is exactly the same as the vision transformer [23] as represented in Fig. 1 of that paper. The multihead attention layer in the transformer encoder has 16 heads, and each head has a dimension of 32. The multilayer perceptron in the transformer block has a hidden dimension of 256 and uses a Gaussian error linear unit (GELU) activation function. The values of these hyperparameters for the transformer encoder were empirically optimized for the best ViT performance.

The structure of the output block differs from that presented in the original ViT [23], by acting on the full output of the final transformer encoder. It contains a one-dimensional global average pooling layer, a fully connected hidden layer with 64 units and a GELU activation function followed by the output layer of the ViT.

The output layer is a fully connected layer with a single output unit with sigmoid activation function. This normalizes the linear output to a probability $\hat{y} \in [0, 1]$ that the input data contains a CW signal.

The sigmoid-normalized output of the ViT is only used in training, while the unnormalized linear output value is used as the learned detection statistic for classification. This follows the approach of previous studies [19,20,33]: the sigmoid-normalized output works best for classification training, but is susceptible to numerical overflow and underflow (latching to 1 or 0, respectively) and is therefore not suitable for use as a detection statistic.

This ViT construction results in a slightly different architecture for each benchmark search type, but the only difference is the input SFT image dimensions, size of the patches (i.e., token dimension), and the number of tokens, while the latent vector size, the transformer encoder with its hyperparameters, and the output block are all unchanged. Thus, by only changing the size (and number) of the patches, we can adapt the same ViT base architecture for different CW search benchmarks.

[1] This table also contains the corresponding values for ViTs for ease of comparison.

### C. Training and validation

We train a ViT network (with architecture described in the previous section) for each of the benchmarks described in Table I. The training dataset for each benchmark contains a large number of signals: 8192 each for the targeted benchmarks and 32 768 each for the directed and all-sky benchmarks, with their signal parameters sampled randomly from their astrophysical ranges given in Table I.

Each ViT is trained on the signals in the training dataset added to independent (Gaussian) noise realizations and an equal number of pure noise inputs. Independent realizations of Gaussian noise are used for every training iteration, which prevents the network from overfitting to features of a particular realization.

The signals used for training are added to the noise at the matched-filtering depth $\mathcal{D}^{90\%}_{\mathcal{F}}$ of the corresponding benchmark given in Table III. The measured ViT detection probability $p_{\mathrm{det}}$ (at fixed false-alarm level of $p_{\mathrm{fa}} = 1\%$) achieved on these input sets can therefore directly be compared to the $\mathcal{F}$-statistic detection probability of $p_{\mathrm{det}} = 90\%$. The signals are added in such a way that the midpoint of the signal bandwidth aligns with the midpoint of the network input, without loss of generality, as the network will slide binwise over the input frequency bins, as discussed in Sec. III D.

Additionally, we also train one all-sky ViT with signals from the full frequency band of 20–1000 Hz, with the remaining search parameters being the same as the all-sky benchmarks in Table I. This allows us to investigate the possibility of training a single ViT for an all-sky search over the full frequency range, instead of having to train separate ViTs for different frequencies. The signals for this wide-frequency ViT are injected at a depth of $\mathcal{D} = 13.4/\sqrt{\mathrm{Hz}}$, corresponding to the average $\mathcal{F}$-statistic sensitivity depth over this frequency range.

For training we use the Adam optimizer [34] with a learning rate of $10^{-4}$ and a batch size of 256. The loss function used is the standard binary cross entropy for classification training, defined as

$$\mathcal{L}(y, \hat{y}) = \frac{1}{N} \sum_{i=1}^{N} [-y^i \log \hat{y}^i - (1 - y^i) \log(1 - \hat{y}^i)], \qquad (3)$$

where $\hat{y}^i \in [0, 1]$ is the normalized output for the $i^{\mathrm{th}}$ input sample, $y^i$ is the corresponding ground-truth label (0 for a noise sample and 1 for a signal sample), and $N$ is the number of inputs in a batch.

During training we also evaluate the ViT on an independent validation dataset at every 10 epochs for the targeted ViTs and at every 100 epochs for the directed and all-sky ViTs. The validation dataset contains the same number of signals independently drawn from the same priors as the training dataset. The loss evaluated on this validation dataset allows us to ensure that the ViT is not overfitting to the signals in the training dataset and thus obtain a more realistic estimate of the performance on unseen data.

The training and evaluation of ViTs was performed on Nvidia A100-SXM4 GPUs with 40 GB of memory. The ViT was implemented in TensorFlow 2 [35] with the Keras API [36]. We used the Weights & Biases platform [37] to monitor training and log losses and metrics during training.

### D. Evaluation metric: Detection probability

We evaluate the ViT sensitivity by computing the detection probability $p_{\mathrm{det}}$ at a fixed false-alarm level of $p_{\mathrm{fa}} = 1\%$, the same as for the $\mathcal{F}$-statistic based searches; see Sec. II B. The ViT is first evaluated on a large number of pure-noise inputs to obtain the noise distribution of the detection statistic, from which we can determine the detection threshold corresponding to $p_{\mathrm{fa}} = 1\%$. The ViT is then evaluated on a large number of inputs with signals added to noise (at fixed depth $\mathcal{D}$), and the resulting $p_{\mathrm{det}}$ is thus obtained as the fraction of signal inputs where the statistic crosses the detection threshold.

In the targeted cases, measuring the network $p_{\mathrm{det}}$ is very fast, as only a single ViT prediction is needed for each noise and signal input, so we compute this for the training dataset at every epoch, and for the validation dataset at every ten epochs. If the estimated $p_{\mathrm{det}}$ (within uncertainties) on the validation dataset crosses a cutoff value of 91%, we stop training early, indicating that the true $p_{\mathrm{det}}$ has reached very close to 90%. The maximum training timespan is one day, as we observed that the $p_{\mathrm{det}}$ is saturated at a value close to 90%.

In the case of the directed and all-sky ViTs, computing $p_{\mathrm{det}}$ is more expensive as we need to cover the 50 mHz bandwidth of the benchmarks. The input bandwidth of our ViTs for all-sky searches is ≈5.6 mHz, and for directed searches it is ≈12.2 mHz, so we slide the ViT (by a single bin at each step) along the frequency axis to fully cover the 50 mHz search bandwidth, using the loudest statistic value obtained for each dataset (the same as for the matched-filter searches described in Sec. II B). For 1800 s SFTs, the frequency resolution of the input data is $1/1800\ \mathrm{s} = 0.56$ mHz, so the ViT has to be evaluated at 90 frequency positions to fully cover the 50 mHz. However, for computing the detection statistic on signal inputs we only need to evaluate the ViT at the known signal position and four neighboring positions (two on each side) for each injection, as this virtually guarantees yielding the loudest statistic over the full frequency band, thereby reducing the cost of computing $p_{\mathrm{det}}$. Because determining the threshold is more expensive in this case, however, we only compute $p_{\mathrm{det}}$ on the validation dataset at every 100 epochs. Contrary to the targeted cases, there is no stopping criterion based on $p_{\mathrm{det}}$ for the directed and all-sky cases. Their training is stopped after three days as the $p_{\mathrm{det}}$ is saturated, and no further improvement is expected.

TABLE IV. Detection probability $p_{\rm det}$ at fixed $p_{\rm fa} = 1\%$ (with 90% confidence interval) achieved by ViTs on the test dataset for a signal population at the matched-filter sensitivity depth $\mathcal{D}^{90\%}_{\mathcal{F}}$ of Table III for each of the search benchmarks.

| $p_{\rm det}$[%] | 20 Hz | 100 Hz | 200 Hz | 500 Hz | 1000 Hz |
|---|---|---|---|---|---|
| Sky-A | $89.6^{+0.5}_{-0.6}$ | $88.9^{+0.6}_{-0.6}$ | $89.7^{+0.5}_{-0.6}$ | $89.6^{+0.5}_{-0.6}$ | $89.6^{+0.5}_{-0.6}$ |
| Sky-B | $89.2^{+0.5}_{-0.6}$ | $89.3^{+0.5}_{-0.6}$ | $89.2^{+0.5}_{-0.6}$ | $88.7^{+0.6}_{-0.6}$ | $87.3^{+0.6}_{-0.6}$ |
| G347 | $86.5^{+0.3}_{-0.3}$ | $87.8^{+0.3}_{-0.3}$ | $86.9^{+0.3}_{-0.3}$ | $86.9^{+0.3}_{-0.3}$ | $86.1^{+0.3}_{-0.3}$ |
| CasA | $88.7^{+0.3}_{-0.3}$ | $88.6^{+0.3}_{-0.3}$ | $88.7^{+0.3}_{-0.3}$ | $87.3^{+0.3}_{-0.3}$ | $84.8^{+0.3}_{-0.3}$ |
| All-sky | $88.3^{+0.3}_{-0.3}$ | $86.1^{+0.3}_{-0.3}$ | $84.7^{+0.3}_{-0.3}$ | $81.5^{+0.4}_{-0.4}$ | $78.2^{+0.4}_{-0.4}$ |

## IV. RESULTS

### A. Performance on a test dataset

Deep neural networks are susceptible to overfitting to the features of the training and validation datasets. In order to estimate their true sensitivity, we evaluate them on an independent test dataset that contains a previously unseen set of signals, drawn from the same priors (Table I). We evaluate the trained ViTs on these signals added to Gaussian noise at $\mathcal{D}^{90\%}_{\mathcal{F}}$ from Table III for the corresponding benchmarks. The resulting $p_{\rm det}$ at a constant $p_{\rm fa} = 1\%$, computed as described in Sec. III D, are given in Table IV for each of the benchmarks. We also compute the 90% upper-limit sensitivity depth $\mathcal{D}^{90\%}_{\rm ViT}$ of the ViTs by measuring $p_{\rm det}$ at different values of $\mathcal{D}$ of the added signals and finding the depth at which it reaches $p_{\rm det} = 90\%$. These sensitivity depths are given in Table III for each of the benchmarks.

For the targeted (10 day) benchmarks, we see that the ViTs achieve near-perfect $p_{\rm det} \approx 90\%$ for almost all cases (except for Sky-B at the highest frequencies), essentially matching an $\mathcal{F}$-statistic search, with no signs of overfitting. Similarly, the corresponding sensitivity depths $\mathcal{D}^{90\%}_{\rm ViT}$ achieved are very close to the matched-filter $\mathcal{D}^{90\%}_{\mathcal{F}}$.

Note that we have previously achieved [20] similar performance on these targeted benchmarks with a CNN network, using similar training times. However, the CNN architecture required substantial manual redesign and tuning away from its "standard" image-classification structure, while the ViT achieves similar performance essentially "out of the box" with no special architecture changes required. This could indicate that the ViT has less restrictive built-in priors about the image morphology, and more naturally satisfies the CW design principles discussed in [20].

For the wide-parameter-space (1 day) benchmarks, the ViT approaches, but does not quite achieve, matched-filtering performance: its $p_{\rm det}$ for signals at $\mathcal{D}^{90\%}_{\mathcal{F}}$ falls short of the WEAVE result by ≈1–5% for the directed benchmarks and by ≈2–11% for the all-sky benchmarks. The corresponding difference in sensitivity depth $\mathcal{D}^{90\%}$ is $\lesssim 1/\sqrt{\rm Hz}$ for the directed benchmarks and $\lesssim 2/\sqrt{\rm Hz}$ for the all-sky benchmarks.

We see that ViT sensitivity declines substantially with increasing signal frequency in the all-sky benchmark, as $p_{\rm det}$ drops by ≈10% from 20 Hz to 1000 Hz. In the directed benchmarks, however, this decline in $p_{\rm det}$ is much less pronounced, and is only noticeable at frequencies of 500 Hz and 1000 Hz for CasA (drop by ≈4%), and only at 1000 Hz for G347 (drop by ≈0.8%).

Increasing frequency affects wide-parameter-space searches in two main ways: (i) the signals get more Doppler shifted and more spread out in the time-frequency plane, and (ii) the number of templates $\mathcal{N}_{\rm T}$ required to cover the parameter space grows. Given that the targeted benchmarks show no substantial drop in performance at higher frequency, this suggests that effect (i) (which is much more pronounced over ten days) does not appear to be a limiting factor for the ViT (or a properly-designed CNN in [20]). Therefore effect (ii), namely the growing number of different signal shapes with time and frequency, is likely the main factor making the problem more difficult for neural networks to learn. Consistent with this explanation we see in Table II that for the all-sky search $\mathcal{N}_{\rm T}$ increases by 4 orders of magnitude in the range 20–1000 Hz, while for the directed searches, $\mathcal{N}_{\rm T}$ increases only by 1 order of magnitude.

We can (approximately) compare the achieved ViT sensitivity on our (one day) directed and all-sky benchmarks with the CNN results presented in [19], albeit using a slightly different timespan of $10^5$ s ≈ 1.16 days. Nevertheless we can compare $p_{\rm det}$, which was defined identically at fixed $p_{\rm fa} = 1\%$ over 50 mHz on signals injected at matched-filtering depth. Comparing our Table IV to Table VI in [19], we see that the ViTs achieve substantially higher $p_{\rm det}$ at every reference frequency.

### B. Generalization in signal strength

The ViTs were trained on a set of signals at fixed sensitivity depth $\mathcal{D}^{90\%}_{\mathcal{F}}$, corresponding to the benchmarks given in Table III. In this section, we study how their sensitivity depends on the strength of the injected test signals. We therefore evaluate the ViTs on sets of signals at different depth added to Gaussian noise and compute $p_{\rm det}$ (at fixed $p_{\rm fa} = 1\%$) using the procedure from Sec. III D. For comparison we also estimate the corresponding $p_{\rm det}$ for an $\mathcal{F}$-statistic search using the sensitivity estimation method of [29].

The resulting $p_{\rm det}$ as a function of depth of the test signals is shown in Fig. 1 for the targeted benchmark of Sky-B at $f = 1000$ Hz, and in Fig. 2 for the all-sky benchmarks at $f = 20$ Hz and $f = 1000$ Hz. These examples include the most challenging targeted and wide-parameter-space benchmark cases for the ViT, where it achieved its worst sensitivity, as seen in Table IV.

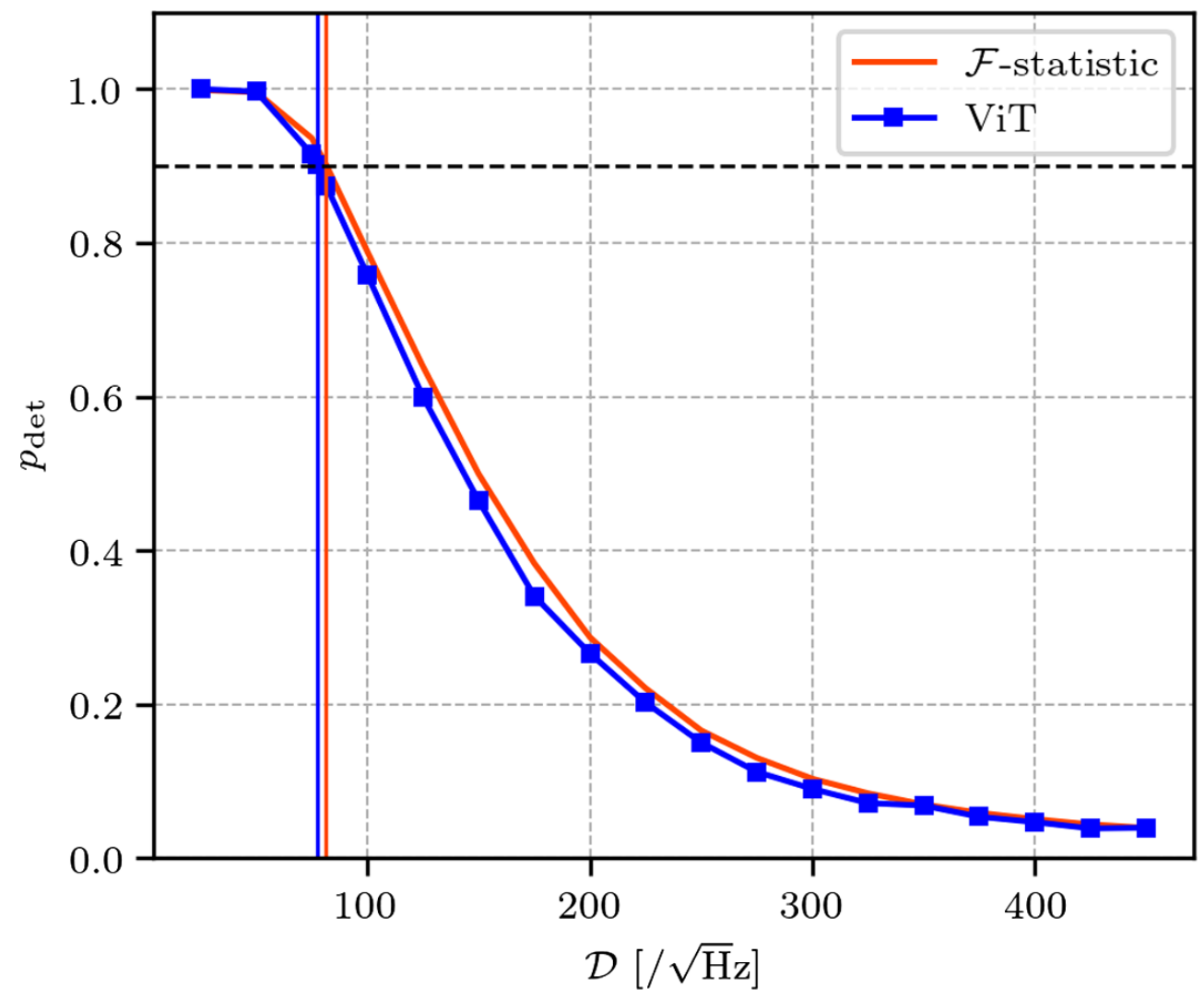

FIG. 1. Detection probability $p_{\rm det}$ of the ViT and the $\mathcal{F}$-statistic, as a function of signal depth $\mathcal{D}$, for the targeted search benchmark of Sky-B at $f = 1000$ Hz.

In all examples shown we see that the ViT sensitivity as a function of signal strength behaves very similar to the matched-filter one, tracking its efficiency curve very closely or at a roughly constant offset. This shows that the ViT generalizes correctly to different signal strengths, despite training at fixed depth $\mathcal{D}^{90\%}_{\mathcal{F}}$, as was previously observed for CNNs [18–21].

### C. Generalization in frequency

Next we study how the performance of the all-sky ViTs (as a representative example) varies for test signals at frequencies different from the training set. For this we create test datasets with frequencies at regular intervals of 20 Hz in the range 20–1000 Hz, with all other parameters drawn from the all-sky benchmark priors of Table I.

In addition to the original five ViTs trained on the five reference frequencies of the benchmarks, we use one additional all-sky ViT trained directly on signals drawn from the full frequency band of 20–1000 Hz. As in Sec. IV A, we test the ViTs at the matched-filter signal depth $\mathcal{D}^{90\%}_{\mathcal{F}}$, linearly interpolated between the measured values at the five $f_{\rm ref}$ of the benchmarks given in Table III.

The resulting detection probability $p_{\rm det}$ for the six all-sky ViTs as a function of frequency of the injected signals is shown in Fig. 3. We see that, as expected, the ViTs show the best $p_{\rm det}$ at the corresponding benchmark frequency they were trained at, while detection probability drops as the offset from the trained $f_{\rm ref}$ increases.

Interestingly, performance seems to drop faster for signal frequencies higher than the trained one compared to signals at lower frequencies. Furthermore, the slope of this drop seems to decrease for ViTs trained at higher frequencies.

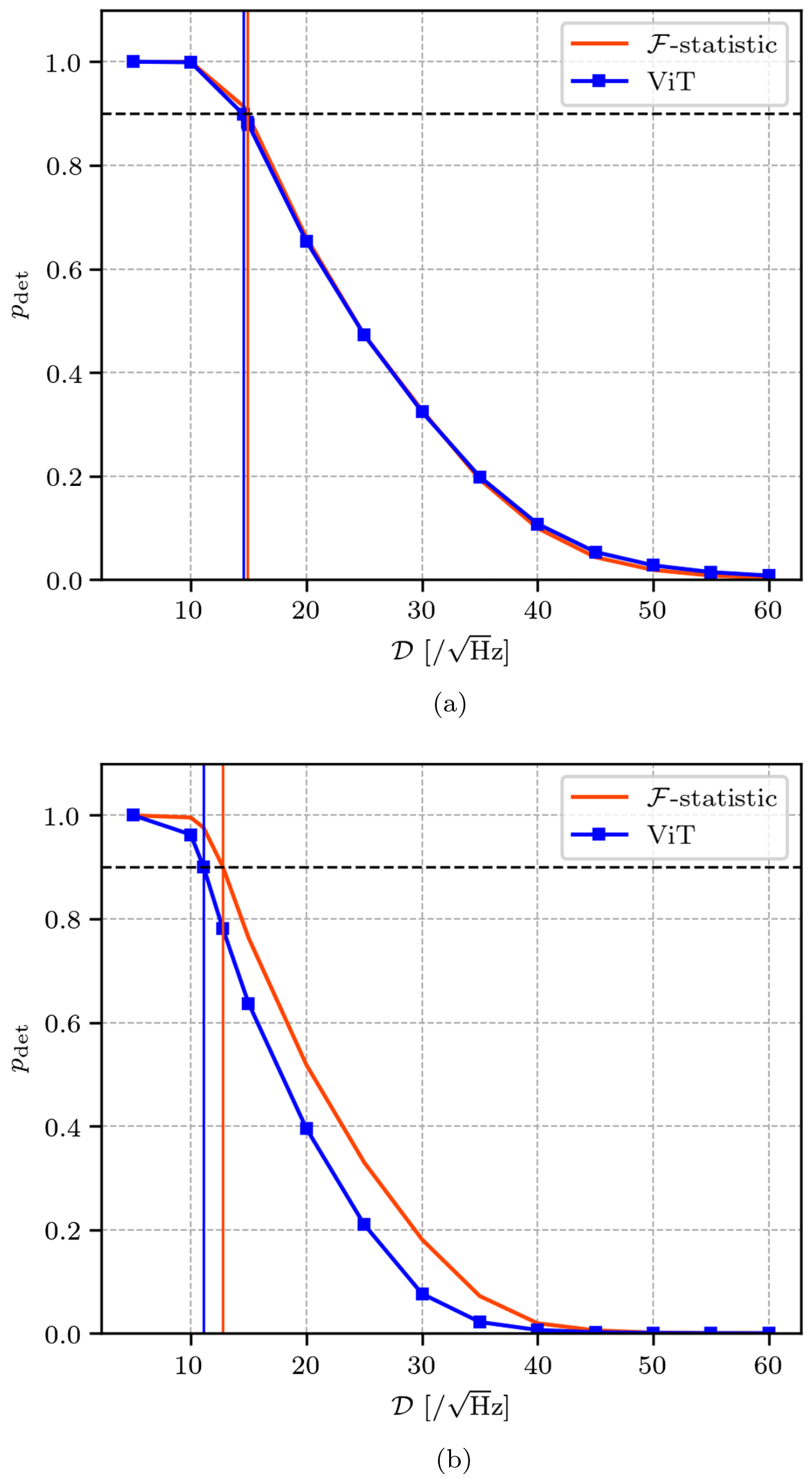

FIG. 2. Detection probability $p_{\rm det}$ of the ViTs and the $\mathcal{F}$-statistic, as a function of signal depth $\mathcal{D}$, for the all-sky search benchmarks at (a) $f = 20$ Hz and (b) $f = 1000$ Hz.

The all-sky ViT trained over the full frequency band 20–1000 Hz performs quite robustly and consistently lies in the range $p_{\rm det} \approx 70$–$80\%$ at all frequencies, with slightly better performance at lower frequencies. Typically for frequencies near the all-sky benchmark frequencies $f_{\rm ref}$, its $p_{\rm det}$ is second only to the ViT trained at that specific frequency. It is remarkable that a network trained on only 32 768 signals for an all-sky search over the full frequency range of 20–1000 Hz can perform with such a high sensitivity. This suggests the practical possibility of training a single ViT for a wide-band search, which can reduce the training cost and logistical hassle of requiring separate ViTs trained for different frequencies.

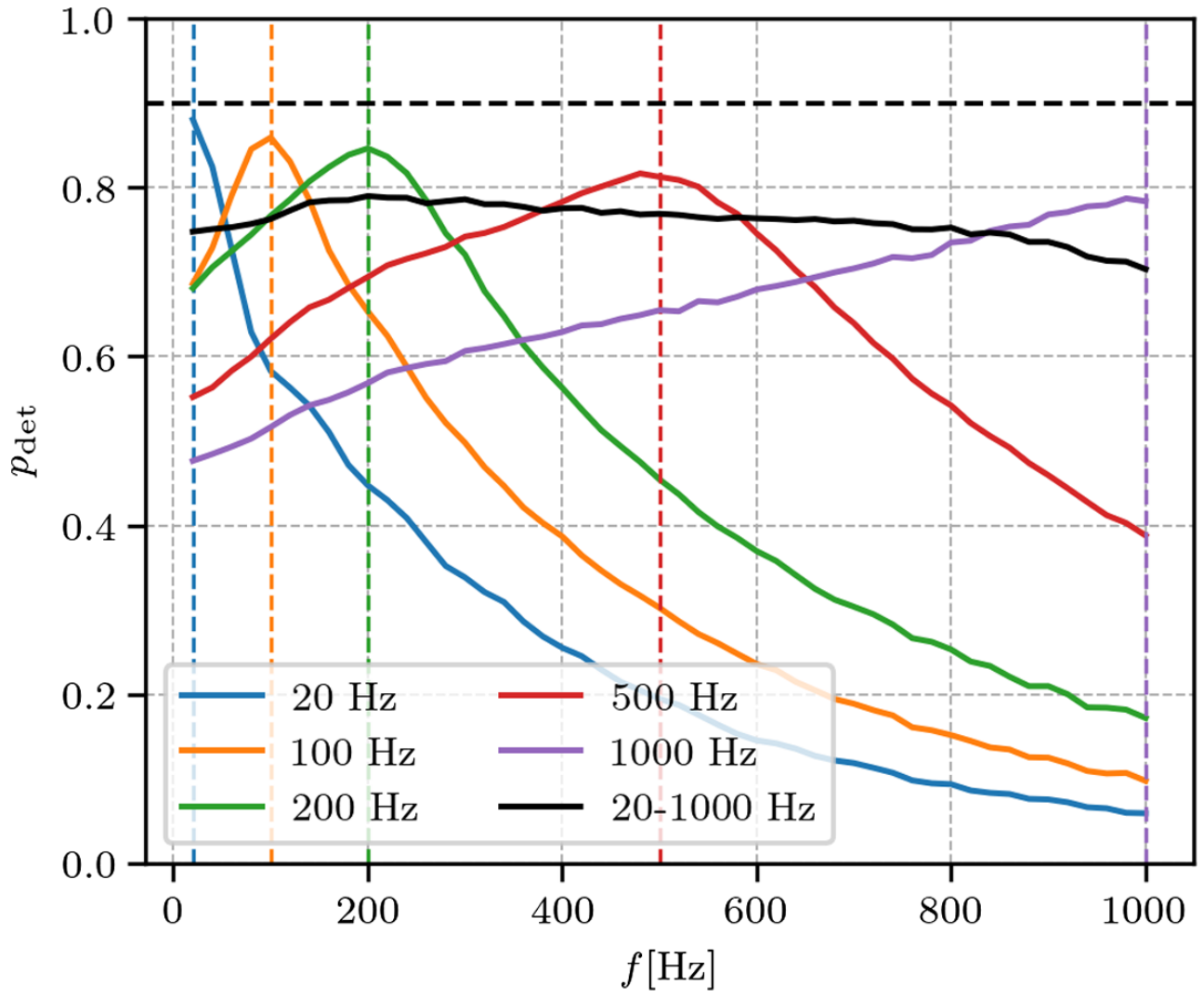

FIG. 3. Detection probability $p_{\rm det}$ (at fixed $p_{\rm fa} = 1\%$ per 50 mHz) as a function of frequency of signals (injected at matched-filter depth $\mathcal{D}^{90\%}_{\mathcal{F}}$) for all-sky ViTs trained at the five benchmark reference frequencies $f_{\rm ref}$ (indicated by vertical dashed lines) as well as ViT trained over all frequencies. The horizontal dashed black line indicates $p_{\rm det} = 90\%$.

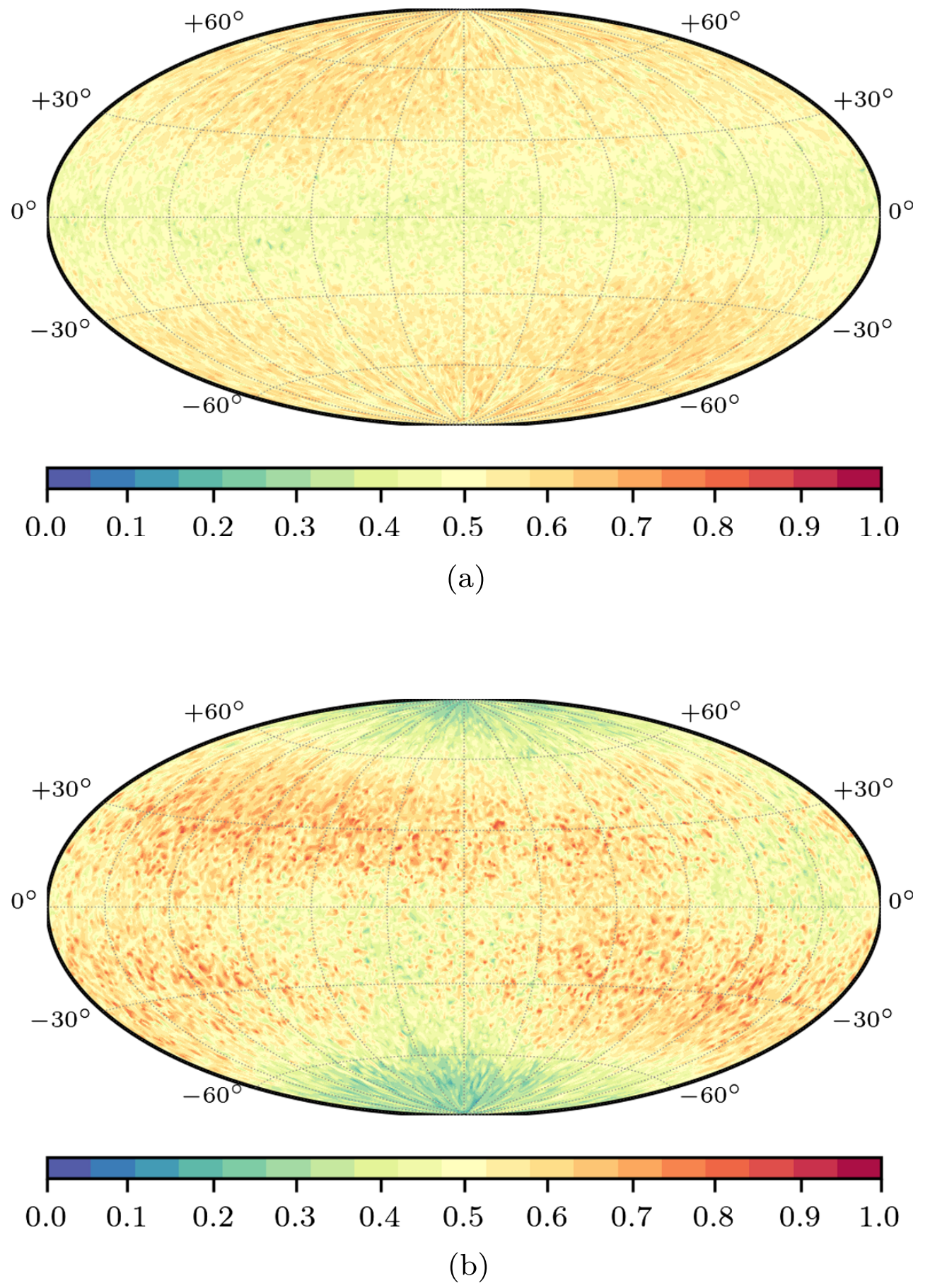

FIG. 4. Detection probability $p_{\rm det}$ as a function of sky position, for signals injected at constant signal power $\rho^2$, chosen for an (average) all-sky $p_{\rm det} = 50\%$. (a) Plot is for the ViT trained and tested at $f_{\rm ref} = 20$ Hz, (b) Plot is for $f_{\rm ref} = 1000$ Hz. Using Hammer projection and equatorial sky coordinates.

### D. Dependence on sky position

It is interesting to test the sensitivity of the all-sky trained ViTs as a function of sky position. In order to do that, we create 32 768 different datasets, each containing 200 signals at a different fixed sky position. The sky position of each dataset is chosen isotropically over the sky. The rest of the signal parameters are sampled from the all-sky benchmark priors of Table I.

Contrary to the constant-depth injections used in our other tests, however, here we add signals to Gaussian noise at a constant signal power $\rho^2$ [defined in Eq. (2)]. The value of $\rho^2$ is chosen such that the all-sky ViT yields an all-sky detection probability of $p_{\rm det} = 50\%$ (at $p_{\rm fa} = 1\%$). Because signal power $\rho^2$ is the only factor affecting the detectability of a signal, independently of sky position (at least for the $\mathcal{F}$-statistic), we would expect the detection probability to be uniform at $p_{\rm det} = 50\%$ over the whole sky.

The measured $p_{\rm det}$ of the all-sky ViTs over the sky is shown in Fig. 4, for frequencies of $f_{\rm ref} = 20$ Hz (top plot) and $f_{\rm ref} = 1000$ Hz (bottom plot), with signals added at a constant $\rho^2 = 39.6$ and 71.9 respectively. For the example at 20 Hz (top plot), we see some slight deviations from the expected mean of $p_{\rm det} = 50\%$, with regions of higher $p_{\rm det}$ around the poles whereas the equatorial band tends to have lower $p_{\rm det}$. In the example at 1000 Hz (bottom plot), we see even more pronounced deviations from the mean $p_{\rm det} = 50\%$, where now the regions of higher $p_{\rm det}$ are more concentrated near the equator, whereas the poles have a lower $p_{\rm det}$. We also see some dependence on right ascension, with a few spots near the equator with higher $p_{\rm det}$ compared to its neighboring region.

It is unclear where these deviations originate, given the training set consisted of isotropically sampled signals over the sky, and the response at fixed $\rho^2$ should ideally be uniform. Similar patterns of deviations have previously been observed for all-sky CNNs as well; see Figs. 4(a) and 4(b) in [21] and 6(e) and 6(f) in [19]. This points to a learned bias in the network sensitivity, a topic that was recently discussed in great detail in [38]. More work is required to understand the origin of these biases in this case, and potential ways to mitigate them, which could result in improved sensitivity.

## V. CONCLUSIONS

We have explored the training and achievable sensitivity of a transformer-based neural-network architecture, namely the vision transformer [23], for the CW search problem.

We have trained ten ViTs on ten-day-timespan targeted searches, and fifteen on one-day-timespan wide-parameter-space searches (ten directed, five all-sky) on a narrow frequency range of 50 mHz at five different reference frequencies, and one all-sky network trained on signals over the full frequency range of 20–1000 Hz. Training times on Nvidia A100 were less than one day for the targeted- and three days for the wide-parameter-space searches.

The targeted ViTs achieved sensitivities equal or very close to that of an $\mathcal{F}$-statistic-based matched filter search. The directed and all-sky ViTs reached record detection probabilities $p_{\mathrm{det}}$ within 2–11% of a WEAVE-based $\mathcal{F}$-statistic search, and within $1\text{–}2/\sqrt{\mathrm{Hz}}$ of its sensitivity depths $\mathcal{D}^{90\%}$, improving on the previous best neural-network sensitivities achieved.

These results show that the standard ViT architecture, without any major changes or redesign, seems well suited for a variety of CW searches. Remarkably, the transformer encoder used for the ten-day-timespan targeted searches and for the one-day-timespan directed and all-sky searches is essentially the same, with the only difference being the size and the number of transformer input tokens, determined by the search timespan and patch size used to cover the input SFT image. In contrast to this, the CNNs used in our previous works [20,21] required CW-specific manual redesign and hyperparameter optimization for every search case to be effective. However, more work is required to establish if ViTs can also beat CNN performance on the harder ten-day-timespan wide-parameter-space search benchmarks studied in [21], which we have postponed to future work.

## ACKNOWLEDGMENTS

This work has utilized the ATLAS computing cluster at the MPI for Gravitational Physics, Hannover, and the HPC system Raven at the Max Planck Computing and Data Facility.

## DATA AVAILABILITY

The data that support the findings of this article are not publicly available. The data are available from the authors upon reasonable request.